\documentclass[journal=aamick,manuscript=article]{achemso}
\usepackage[version=4]{mhchem} 
\usepackage{tabularray}
\usepackage{amsmath}
\usepackage{bm}
\usepackage{epsfig}
\usepackage{epstopdf}
\usepackage{multirow}
\usepackage{graphicx}
\usepackage{subcaption}
\usepackage{siunitx}

\author{Vuong Van Thanh}
\email{thanh.vuongvan@hust.edu.vn}
\affiliation{School of Mechanical Engineering, Hanoi University of Science and Technology, Hanoi 100000, Viet Nam}
\author{Nguyen Tuan Hung}
\email{nguyen.tuan.hung.e4@tohoku.ac.jp}
\affiliation{Frontier Research Institute for Interdisciplinary Sciences, Tohoku University, Sendai, 980-8578 Japan}

\title[title]{Strain Effect on Rashba Splitting and Phonon Scattering to Improve Thermoelectric Performance of 2D Heterobilayer MoTe$_{2}$/PtS$_{2}$}

\abbreviations{}

\begin{document}

\begin{abstract}
Rashba spin-orbit coupling significantly modifies the electronic band structure in two-dimensional (2D) van der Waals (vdW) heterobilayers, which may enhance their thermoelectric (TE) properties. In this study, we use first-principles calculations and Boltzmann transport theory to explore the strain effect on the TE performance of the 2D vdW heterobilayer MoTe$_{2}$/PtS$_{2}$. A strong Rashba spin-splitting is observed in the valence band, resulting in an increase in the Seebeck coefficient for p-type. The lattice thermal conductivity of MoTe$_{2}$/PtS$_{2}$ is remarkably low about of 0.6 Wm$^{-1}$K$^{-1}$ at $T = 300$ K due to large anharmonic scattering. Furthermore, biaxial strain enhances the power factor (PF) by introducing band convergence. At a strain of 2\%, the optimal PF for the n-type material reaches \SI{170}{\micro\watt}/cmK$^{2}$, indicating approximately 84.78\% increase compared to the unstrained state (\SI{92}{\micro\watt}/cmK$^{2}$). Given the low lattice thermal conductivity, the optimized figure of merit $ZT$ achieves up to 0.88 at 900 K for n-type. Our findings indicate that MoTe$_{2}$/PtS$_{2}$ is a highly promising candidate for 2D heterobilayer TE materials, owing to its strong Rashba splitting and significant anharmonicity.
\end{abstract}

\section{Introduction}
Nowadays, fossil energy sources are causing environmental pollution and are gradually depleting due to the increasing energy demand. 
Therefore, a thermoelectric (TE) generator, which can directly convert waste heat into electricity, is a practical solution to address the issue of fossil energy~\cite{Biswas2022,Heremans2013,Vining2009}.
To evaluate the performance of the energy conversion of the TE device, the figure of merit, $ZT = S^{2}\sigma T/(\kappa_\text{el} + \kappa_\text{ph})$, is used, where $S$, $\sigma$, $\kappa_\text{el}$, $\kappa_\text{ph}$, and $T$ are the Seebeck coefficient, electrical conductivity, thermal conductivities by electron, thermal conductivities by lattice, and temperature, respectively~\cite{Goldsmid2010}.
Here, the $S^{2}\sigma$ is the TE power factor (PF). 
A material considered suitable for the TE device should have a high $ZT$ or PF~\cite{van2024janus}. 


Several methods have been proposed to enhance the PF and $ZT$, such as resonant state~\cite{kaidanov1992resonant}, chemical material~\cite{zhao2020chemical}, quantum enhancement effect in low-dimensional semiconductors~\cite{Hicks1993,Dresselhaus2007,hung2016quantum,hung2021origin}, band convergence~\cite{tang2015convergence,hung2019thermoelectric}, and Rashba spin-orbit coupling (SOC)~\cite{yuan2018one}. Among them, the Rashba SOC depends only on the intrinsic properties of the material without requiring additional external factors. The Rashba SOC leads to the enhanced electronic density of states, resulting in a larger Seebeck coefficient~\cite{xia2024thermoelectric}. Yuan \textit{\textit{et al.}}~\cite{yuan2018one} indicated that Rashba SOC can enhance the TE performance by 78\% for BiSb monolayer. They point out that Rashba SOC not only induces a one-dimensional (1D) density of states near the band edge but also improves the carrier lifetimes in the Rashba spin-splitting system, leading to a significant enhancement of the PF. 
Bera \textit{\textit{et al.}}~\cite{bera2021spin} studied the effect of Rashba SOC on the TE properties of HfX${_2}$ (X= S, Se) and Janus HfSSe monolayers, and they indicated that Rashba SOC enhances the PF for n-type carriers of these materials. The highest achieved $ZT$ values of HfS${_2}$, HfSe${_2}$, and HfSSe monolayers are 0.90, 0.84, and 0.81 at 600 K, respectively. Some recent studies also demonstrated that the Rashba SOC can play a crucial role in enhancing the TE properties of monolayer materials~\cite{wang2018nontrivial,duan2020improved}. However, monolayer materials with strong Rashba SOC often exhibit high thermal conductivity. For example, $\kappa_\text{ph}$ of Si$_{2}$AsSb and Si$_{2}$AsSb is 72.2 W/mK and 49.3 W/mK at room temperature, respectively~\cite{xia2024thermoelectric}, which is larger than that of well-known TE materials, such as Bi$_{2}$Te$_{3}$ ($\sim 1.4$ W/mK)~\cite{qiu2009molecular}, and PbTe ($ \sim  2.0$ W/mK)~\cite{pei2011high}. Thus, searching for two-dimensional (2D) materials with high Rashba SOC to enhance PF, along with low $\kappa_\text{ph}$, is essential to achieve a high $ZT$. 

The 2D van der Waals (vdW) heterobilayers could be a potential candidate for 2D materials with both high Rashba SOC and low $\kappa_\text{ph}$. Gupta \textit{\textit{et al.}}~\cite{gupta2021dictates} found that the 2D vdW heterobilayer MoTe$_{2}$/PtS$_{2}$ with AB stacking structure has a strong Rashba SOC~\cite{gupta2021dictates}, in which the Rashba parameter, $\alpha_{R}$ $\sim $ 8 eV\AA, is much larger than that of other 2D materials, such as Janus WSeTe ($\alpha_{R}$ = 0.92 eV\AA)~\cite{yao2017manipulation} and BiTeI monolayers ($\alpha_{R}$ = 1.86 eV\AA)~\cite{ma2014emergence}. In addition, it was suggested that forming vdW heterobilayers can effectively reduce the lattice thermal conductivity~\cite{noshin2018thermal,jia2019excellent,mohanta2019superhigh}. 
Hu \textit{et al.}~\cite{hu2020surprisingly} reported that $\kappa_\text{ph}$ of the 2D vdW heterobilayer black-phosphorus/blue-phosphorus at $T$ = 300 K is 3.85 and 4.18 W/mK along the $x$- and $y$-directions, respectively. These values are significantly lower than those of black-phosphorus monolayer (7.49 and 23.43 W/mK) and blue-phosphorus monolayer (21.76 and 21.76 W/mK), respectively. 
Thus, it suggest that heterobilayer MoTe$_{2}$/PtS$_{2}$ could has low lattice thermal conductivity. Moreover, in a previous theoretical study, Yin \textit{\textit{et al.}}~\cite{yin2022type} indicated that the energy band gap of MoTe$_{2}$/PtS$_{2}$ is 1.26 eV by using the hybrid functional HSE calculation, which is a suitable band gap for TE materials. On the other hand, the 2D vdW heterobilayer MoSe$_{2}$/PtSe$_{2}$, which has a similar structure of MoTe$_{2}$/PtS$_{2}$, has also been successfully synthesized experimentally~\cite{wang2023enhanced}.
A combination of large Rashba SOC, a suitable band gap, and possible low $\kappa_\text{ph}$ and forming heterobilayer suggests that 2D vdW heterobilayers MoTe$_{2}$/PtS$_{2}$ could be a promising candidate for 2D TE material.

In this study, we investigate the effect of strain on the electron, phonon, and TE performance of 2D vdW heterobilayer MoTe$_{2}$/PtS$_{2}$ by using first-principles calculations and Boltzmann transport theory. The strain promotes the convergence of multivalley bands, resulting in an enhancement of both the PF and $ZT$. We find that the band convergence can be achieved for both n-type and p-type by a biaxial strain of $\varepsilon = 0.02$ and $\varepsilon = 0.03$, respectively. The maximum PF of n-type MoTe$_{2}$/PtS$_{2}$ reaches \SI{170}{\micro\watt}/cmK$^{2}$ W/mK$^{2}$ at $T$ = 300 K. The lattice thermal conductivity of MoTe$_{2}$/PtS$_{2}$ is remarkably low about of 0.6 W/mK at room temperature) because of strong phonon scattering, and the maximum $ZT$ is 0.88 at 900 K for n-type. 

\section{Methodology}
We use density functional theory (DFT) with the Quantum ESPRESSO package~\cite{giannozzi2009,nguyen2022QE} to perform the calculations. The optimized norm-conserving Vanderbilt pseudopotentials~\cite{hamann2013optimized} are adopted to calculate the exchange-correlation functionals~\cite{perdew1996generalized}, in which we use the Perdew-Burke-Ernzerhof (PBE) functionals. 
The Heyd-Scuseria-Ernzerhof (HSE) method~\cite{heyd2003hybrid} is used to accurately calculate the band gap. The Wannier90 package~\cite{mostofi2014updated} is used to interpolate the band dispersion from the HSE calculation. The SOC is considered in our calculations. An energy cutoff of 80 Ry is used for the plane wave basis set. To eliminate interactions due to periodic boundary conditions, a vacuum region of 20 {\AA} is set along the $z$-direction, perpendicular to the sheet plane. The phonon dispersion is calculated by using density-functional perturbation theory (DFPT)~\cite{baroni2001phonons} with a $\bm{q}$-point grid of $8 \times 8 \times 1$, which is selected based on the convergence test. The lattice constants and atomic positions are optimized via the BFGS algorithm~\cite{nguyen2022QE} with convergence criteria of 0.00001 Ry/a.u. and  0.05 GPa for atomic forces and stress components, respectively. The Grimme-D2 method is used to describe the van der Waals (vdW) interaction between the MoTe$_2$ and PtS$_2$ layers~\cite{grimme2016dispersion}. For the optimization of the structure and the calculation of electronic properties, the $\bm{k}$-mesh is set to $12 \times 12 \times 1$. The $ab$ $initio$ molecular dynamics (AIMD) simulations are performed to evaluate thermal stability. A $3\times 3 \times 1$ supercell is employed within the NVT ensemble, with temperature control regulated by the Nose-Hoover method~\cite{Martyna1992} and a time step of 1 fs.

The Seebeck coefficient $S$, electrical conductivity $\sigma$, and electronic thermal conductivity $\kappa_\text{el}$ are calculated by solving the Boltzmann transport theory for electrons by using the BoltzWann code~\cite{pizzi2014boltzwann}. The deformation potential approximation is applied to estimate the relaxation time constant for the Boltzmann transport theory, which is estimated by both longitudinal acoustic (LA) and longitudinal optical (LO) phonon modes in the present study. The lattice thermal conductivity $\kappa_\text{ph}$ is calculated by solving the Boltzmann transport equation for phonons by using the ShengBTE package~\cite{Li2014}. The anharmonic interatomic force constants are calculated using the THIRDORDER.py tool~\cite{Li2014} with the $3\times 3 \times 1$ supercell. A dense grid of $48 \times 48 \times 1$ is used to evaluate the Boltzmann transport equation for phonons. The Bader code~\cite{tang2009grid} is used to analyze the charge transfer. The parameters in the present calculations, including cutoff energy, $\bm k$-points, and $\bm q$-points, are carefully selected based on convergence testing.

\section{Results and discussion}
\subsection{Optimized structure and structural stability}

\begin{figure*}[t]
  \centering \includegraphics[clip,width=12cm]{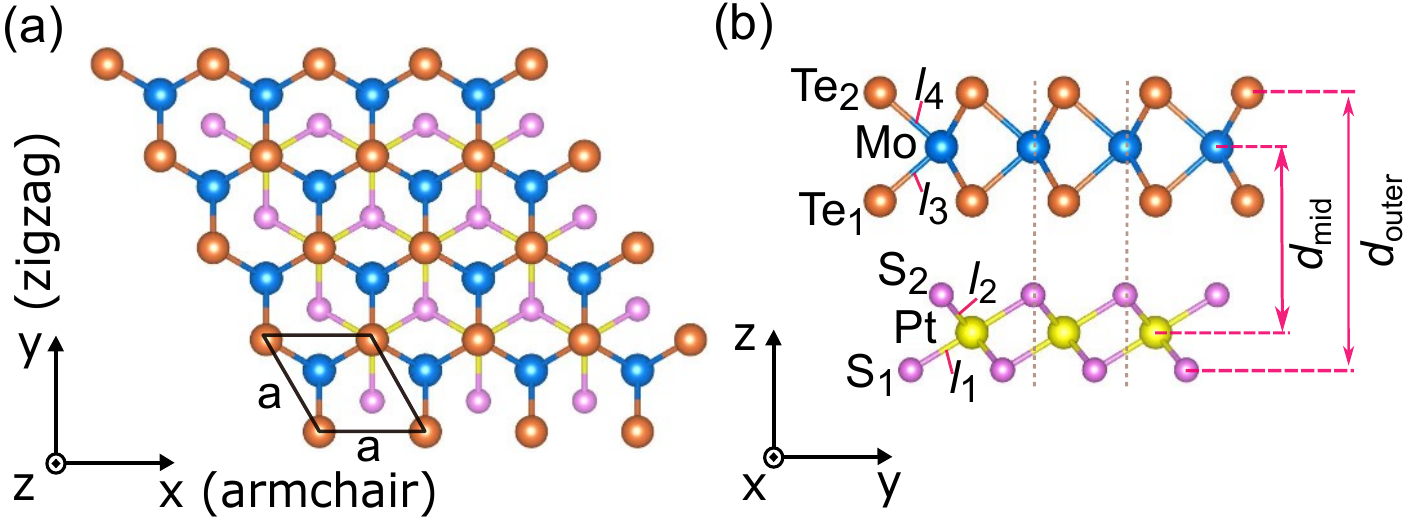}
  \caption{Top (a) and side (b) views of the 2D vdW heterobilayer MoTe$_{2}$/PtS$_{2}$, in which the black box indicates the unit cell, and $a$, $l_{1,2,3,4}$, $d_{\text{mid}}$, and $d_{\text{outer}}$ denote the lattice constant, bond lengths, interlayer distance, and distance between the two outermost atoms, respectively.}
  \label{fig:model}
\end{figure*}

In Figures~\ref{fig:model}(a) and (b), we show the top and side views of the optimized structure of the 2D vdW heterobilayer MoTe$_{2}$/PtS$_{2}$ with AB stacking order, respectively. In the AB stacking, the S atom from the PtS$_{2}$ layer is aligned with the Mo atom from the MoTe$_{2}$ layer, as shown in Figure~\ref{fig:model}(b). The optimized lattice constant $a$, interlayer distance $d_{\text{mid}}$ and outer distance $d_{\text{outer}}$ of the heterobilayer are 3.560 {\AA}, 6.062 {\AA} and 9.096 {\AA}, respectively. We note that the optimized lattice constants of the MoTe$_{2}$ and PtS$_{2}$ monolayers are 3.550 {\AA} and 3.580 {\AA}, respectively. Thus, the lattice mismatch between MoTe$_{2}$ and PtS$_{2}$ is small ($\sim0.85\%$), which suggests that heterobilayer MoTe$_{2}$/PtS$_{2}$ can be formed without bending strain or moir\'e patterns. The Pt-S$_{1}$, Pt-S$_{2}$, Mo-Te$_{1}$ and Mo-Te$_{2}$ bond lengths (see Figure~\ref{fig:model}(b)) are $l_{1}$ = 2.404 {\AA}, $l_{2}$ = 2.396 {\AA}, $l_{3}$ = 2.735 {\AA}, and $l_{4}$ = 2.737 {\AA}, respectively, which are in good agreement with previous report~\cite{almayyali2020stacking}.

To examine the stability of the 2D vdW heterobilayer MoTe$_{2}$/PtS$_{2}$, we first calculate the binding energy $E_{b}$: 
\begin{equation}
  \label{eq:binding}
     E_{b} = E_{\mathrm{MoTe_2/PtS_2}}-E_\mathrm{MoTe{_2}} - E_\mathrm{PtS{_2}},
\end{equation}
where $E_\mathrm{MoTe{_2}/PtS{_2}}$, $E_\mathrm{MoTe{_2}}$, and $E_\mathrm{PtS{_2}}$ are the total energy of the heterobilayer, MoTe${_2}$, and PtS${_2}$ monolayers, respectively. The calculated $E_b = -0.36$ eV is consistent with the previous report (-0.33 eV)~\cite{yin2022type}. Moreover, Almayyalia \textit{et al.}~\cite{almayyali2020stacking} also indicated that the AB stacking has the lowest binding energy compared to that of the AA, AA$^\prime$, and AB stacking orders. Therefore, the AB stacking is chosen to investigate the TE properties of MoTe$_{2}$/PtS$_{2}$ in the present study. 

Next, to examine the dynamic stability of MoTe${_2}$/PtS${_2}$, we calculate the phonon dispersion. As shown in Figure~\ref{fig:k_pp-phonon}(a), MoTe${_2}$/PtS${_2}$ exhibits good dynamic stability, as its phonon dispersion contains no imaginary frequencies. In additional, the calculated the elastic constants $C_{11} = 184.04$ N/m, $|C_{12}| = 51.94$ N/m, and $C_{66} = 66.5$ N/m, satisfy the Born criterion for mechanical stability ($C_{11} > |C_{12}| > 0$ and $C_{66} > 0$)~\cite{mouhat2014necessary}. Thus, MoTe${_2}$/PtS${_2}$ is mechanical stable. We also perform the AIMD simulations at 900 K, as shown in Figure S1 in Supporting Information, to evaluate the thermal stability. The obtained results reveal that the energy fluctuation is small, and the atomic configuration remains stable, showing no obvious distortion. This implies that MoTe$_{2}$/PtS$_{2}$ exhibits strong thermal stability and is well-suited for high-temperature TE applications.

\subsection{Energy band structure and carrier mobility}

\begin{figure}[t]   
  \centering \includegraphics[clip,width=9cm]{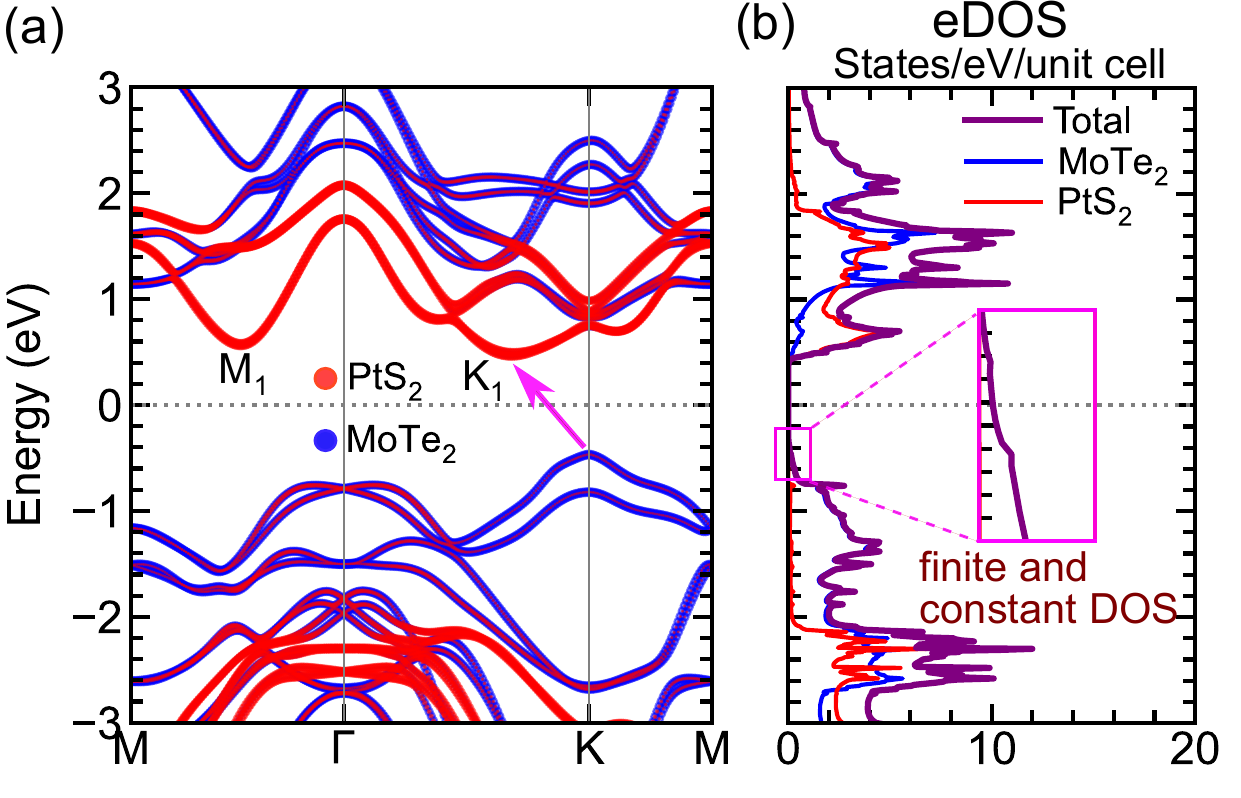}
 \caption{(a) Energy band structure along the high symmetry points and (b) electronic density of states (eDOS) of the 2D vdW heterobilayer MoTe$_{2}$/PtS$_{2}$ using the HSE + SOC method. The blue and red colors indicate the contribution of the MoTe$_{2}$ and PtS$_{2}$ layers, respectively.}
\label{fig:band_pdos}
\end{figure}

In Figure~\ref{fig:band_pdos}(a), we show the band structure of MoTe$_{2}$/PtS$_{2}$ by using HSE + SOC calculations. Our calculations show that MoTe$_{2}$/PtS$_{2}$ exhibits an indirect band gap, with the conduction band minimum (CBM) located at K$_{1}$ (along the $\Gamma$-K direction) and the valence band maximum (VBM) located at the K point. The calculated band gap of MoTe$_{2}$/PtS$_{2}$ is 0.93 eV, which is consistent with previous theoretical reports (1.06 eV)~\cite{yin2022type}. 

In Figure~\ref{fig:band_pdos}(b), we calculate the projected electronic density-of-states (eDOS) to examine the contribution of each material layer to the overall band structure. The calculated results indicate that the CBM is mainly contributed by the PtS$_{2}$ layer, while the VBM mainly originates from the MoTe$_{2}$ layer. As shown in Figure~\ref{fig:band_pdos}(a), the Rashba splitting and a VBM in the valence band are found near the $\Gamma$ point ($-0.8$ eV) and at the K point ($-0.5$ eV), respectively. This leads to one-dimensional (1D) van Hove singularity and a constant 2D density of states around $-0.8$ eV and $-0.5$ eV~\cite{Hung2017}, respectively, as shown in Figure~\ref{fig:band_pdos}(b). These eDOS characteristics could make the outstanding TE properties of 2D materials, such as Janus $\gamma$-Ge$_{2}$SSe~\cite{Thanh2023} and InSe~\cite{hung2019thermoelectric,Hung2017}. Thus, we also expect that the 2D vdW MoTe$_{2}$/PtS$_{2}$ heterobilayer will show high TE performance. 

\begin{figure*}[t]
\centering \includegraphics[clip,width=16cm]{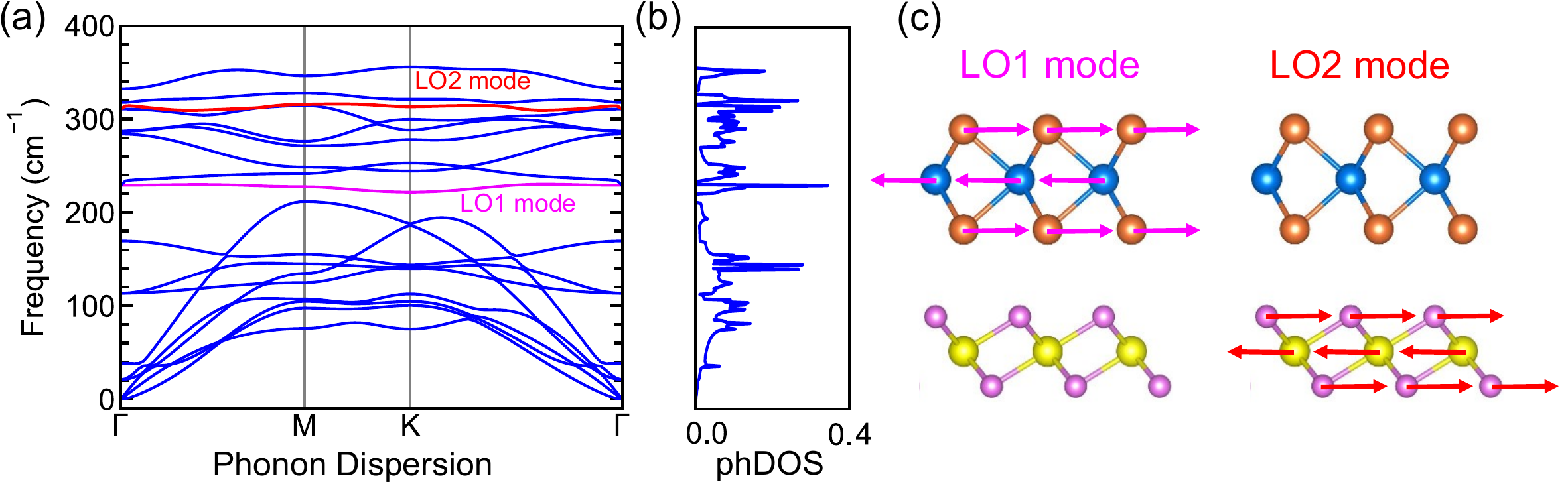}
\caption{(a) Phonon dispersion and (b) phonon density of states (phDOS) of 2D vdW heterobilayer MoTe$_{2}$/PtS$_{2}$. (c) Side view of the atomic vibrations of the longitudinal optical phonon modes at the $\Gamma$ point, in which LO$_{1}$ and LO$_{2}$ phonon modes are found on the MoTe$_2$ and PtS$_2$ layers, respectively.}
\label{fig:k_pp-phonon}
\end{figure*}

\begin{table*}[t]
\centering
\caption{Effective masses $m^{*}/m_0$, where $m_0$ is the mass of free electron, deformation potential $E_{d}$ (eV), elastic modulus $C_\text{2D}$ (N/m), carrier mobilities contributed by LO1 ($\mu_0^\text{LO1}$), LO2 ($\mu_0^\text{LO2}$), LA modes ($\mu_0^\text{LA}$) and combined of these modes ($\mu_0^\text{total}$) (cm$^{2}$V$^{-1}$s$^{-1}$), and total relaxation time $\tau^\text{el}$ ($\times  10^{-14}$ s) of the 2D vdW heterobilayer MoTe$_{2}$/PtS$_{2}$ at 300 K.}
\begin{tabular}{c c c c c c c c c c}
\hline\hline
Direction & Carrier & $m^*/m_0$ & $E_d$ & $C_\text{2D}$ & $\mu_{0}^\text{LO1}$ & $\mu_{0}^\text{LO2}$ & $\mu_{0}^\text{LA}$ & $\mu_{0}^\text{total}$ & $\tau^\text{el}$\\\hline

\multirow{2}{*}{armchair} & e & 1.344 & 1.23 & \multirow{2}{*}{184.04} & 34.58 & 99.38 & 1434.16 & 25.20  & 1.926 \\  
                          & h & -0.761 & -1.21 &  & 61.07 & 175.51 & 4634.58 & 44.87 & 1.942 \\ 
\multirow{2}{*}{zigzag} & e & 1.344 & -1.27 & \multirow{2}{*}{184.04} & 34.58 & 99.38 & 1345.24 & 25.17 & 1.924 \\ 
                        & h & -0.757 & -1.21 &  & 61.40 & 176.44 & 4659.07 & 45.11 & 1.942 \\ 
\hline\hline
\end{tabular}
\label{tab:2}
\end{table*}

Since carrier mobility plays a crucial role in TE properties, we next determine the carrier mobility of MoTe$_{2}$/PtS$_{2}$. For 2D systems, the deformation potential approximation (DPA) is widely used to calculate mobility~\cite {Thanh2023,yuan2023first,luo2021pd4s3se3}. However, conventional DPA accounts only for electron-phonon scattering by longitudinal acoustic (LA) phonons. Thus, the DPA often overestimates the carrier mobility values by ignoring other scattering.
For example, in a previous theoretical study, Cheng and Liu~\cite{cheng2018limits} reported that the electron mobility of H-MoTe$_2$ is 37 cm$^{2}$V$^{-1}$s$^{-1}$ by considering both LA and longitudinal optical (LO) phonon scattering, whereas the DPA method predicts a value of 338 cm$^{2}$V$^{-1}$s$^{-1}$. It highlights the significant contribution of LO phonon scattering to carrier mobility, particularly in the 2D polar system~\cite{bao2025rational,hung2018two}. Thus, in the present study, we also considered the influence of LO phonons when calculating the carrier mobility of the 2D vdW heterobilayer MoTe$_{2}$/PtS$_{2}$. 

The carrier mobilities $\mu_0^\text{LA}$ and $\mu_0^\text{LO}$ for the 2D material are given by~\cite{Thanh2023,yuan2023first,luo2021pd4s3se3,bao2025rational}: 
\begin{equation}
  \label{eq:mu_LA}
   \mu_0^\text{LA} = \frac{e\hbar^{3}C_\text{2D}}{k_{B}Tm_{x(y)}^{*}\sqrt{m_{x}^{*}m_{y}^{*}}E_{d}^{2}},
\end{equation}
and
\begin{equation}
  \label{eq:mobility_LO}
    \mu_0^\text{LO} = \frac{M_{A}M_{B}S_{0}d_{0}^{2}(\epsilon\epsilon_0)^{2}(\hbar \omega)^{2}}{e^{3}\hbar n_{\hbar \omega}m_{x(y)}^{*}} \frac{1}{Z_{A}^{2}M_{B} 
    + 4Z_{B}^{2}M_{A} 
    + 4\sqrt{M_{A}M_{B}} \left| Z_{A}Z_{B} \right|},
\end{equation}
for the LA and LO phonon scattering, respectively, where $e$, $\hbar$, $k_{B}$, and $T$ are the elementary charge, reduced Planck constant, Boltzmann constant, and temperature, respectively. $d_0$, $\epsilon$, and $\epsilon_{0}$ represent effective thickness, in-plane dielectric constant, and vacuum permittivity, respectively. Here, $d_{0}$ is determined using the following expression: $d_{0} = d_{\text{outer}} + r_{\text{S}} + r_{\text{Se}}$, where $r_{\text{S}}$ and $r_{\text{Se}}$ are the vdW radii of S and Se atoms, respectively~\cite{alvarez2013cartography}. $M_{A}$ and $M_{B}$ are the atomic mass, and $Z_{A}$ and $Z_{B}$ are the Born effective charges of metal $A$ and chalcogen $B$ atoms, respectively (see Table S1 in Supporting Information). $n_{\hbar \omega}$ is the Bose-Einstein distribution at the LO phonon frequency at the $\Gamma$ point. $m_{x}^{*}$ and $m_{y}^{*}$ are the effective masses along the $x$- (armchair) and $y$-directions (zigzag). To calculate $m_x^*$ and $m_y^*$, we use a $\sqrt{3} \times 1$ supercell (see Figure S2 in Supporting Information). $C_\text{2D}$ denotes the elastic constant and is defined as $C_\text{2D}=[\partial^{2}E_{\text{total}}/\partial(\Delta l/l_{0})^2]/S_{0}$, where $S_{0}$ is the equilibrium unit cell area, $E_{\text{total}}$ is the total energy after applying a uniaxial strain, $l_{0}$ is the lattice constant, $\Delta l = (l-l_{0})/l_{0}$) is the corresponding lattice distortion (see Figure S3 in Supporting Information). $E_{d}$ is the deformation potential constant and can be calculated as $E_{d} = \Delta E/\varepsilon$ where $\Delta E$ is the shift in the energy level of the CBM (or the VBM) relative to the vacuum level (see Figure S4 in Supporting Information). 

The 2D vdW heterobilayer MoTe$_{2}$/PtS$_{2}$ has two distinct LO phonon modes, LO1 and LO2, correspond to the vibrations of the MoTe$_{2}$ and PtS$_{2}$ layers, respectively, as shown in Figure~\ref{fig:k_pp-phonon}(c). Therefore, the total mobility $\mu_0^\text{total}$ can be obtained by using Matthiessen’s rule as follows~\cite{cheng2018limits}:
\begin{equation}
  \label{eq:mobility_total}
    \frac{1}{\mu_0^\text{total}}=\frac{1}{\mu_0^\text{LO1}}+\frac{1}{\mu_0^\text{LO2}} + \frac{1}{\mu_0^\text{LA}}.
\end{equation}
The calculated values of $\mu_0^\text{LO1}$, $\mu_0^\text{LO2}$ and $\mu_0^\text{total}$ of MoTe$_{2}$/PtS$_{2}$ without strain at $T$ = 300 K are listed in Table~\ref{tab:2}. 

Our results indicate that LO phonon scattering significantly limits the mobility of the 2D vdW heterobilayer MoTe$_{2}$/PtS$_{2}$, resulting in lower carrier mobility compared to the values predicted using DPA alone, as listed in Table~\ref{tab:2}. For example, the hole mobility $\mu_0^\text{LO1}$ is 61.07 (or 61.40) cm$^{2}$V$^{-1}$s$^{-1}$, which leads to reduce two orders of magnitude of $\mu_0^\text{total}$ about 44.87 (or 45.11) cm$^{2}$V$^{-1}$s$^{-1}$ compared with $\mu_0^\text{LA}$ about 4634.58 (or 4659.07) cm$^{2}$V$^{-1}$s$^{-1}$) in the armchair (or zigzag) direction. This is because of the large Born effective charge of Mo ($Z_\text{Mo}=-2.2e$) and Te ($Z_\text{Te}=1.1e$) atoms of the MoTe$_2$ layer, which is much larger than that of the WSe$_2$ layer ($Z_\text{W}=-0.57e$ and $Z_\text{Se}=0.27e$~\cite{bao2025rational}) in the Janus heterobilayer WSe$_2$/SWSe. Since the strength of the LO phonon scattering is closely related to the magnitude of the Born effective charge. Thus, a large Born effective charge correlates with stronger carrier scattering, leading to reduced mobility in Eq.~\eqref{eq:mobility_LO}. Nevertheless, the Born effective charge reflects strong polarizability, which is the cause of a strong Rashba effect~\cite{gupta2021dictates}. The total relaxation time $\tau^\text{el}=m^*\mu_0^\text{total}/|e|$ is also listed in Table~\ref{tab:2}. The relationship between temperature and the relaxation time $\tau^{el}$ at different strain values is shown in Figures S5 to S8 (Supporting Information).

\subsection{Mechanical properties}
When applying the 2D vdW heterobilayer MoTe$_{2}$/PtS$_{2}$ in flexible devices, it is essential to consider both the mechanical properties and the thermoelectric (TE) performance under strain. The mechanical behavior of 2D materials in the elastic regime can be characterized by Young's modulus, $Y_{\text{2D}}(\theta)$, and Poisson's ratio, $\upsilon(\theta)$, which is determined using the following equations~\cite{van2024janus,Thanh2023,hung2018two}:
\begin{equation}
  \label{eq:young}
Y_{\text{2D}}(\theta) = \frac{C_{11}C_{22}-C_{12}^{2}}{C_{11}\sin^4(\theta) + C_{22}\cos^4(\theta) - \left(2C_{12} - \frac{C_{11}C_{22}-C_{12}^{2}}{C_{66}}\right) \sin^2(\theta) \cos^2(\theta)},
\end{equation}
and 
\begin{equation}
  \label{eq:poisson}
\upsilon(\theta) = \frac{C_{12}(\sin^4(\theta)+\cos^4(\theta))-\left(C_{11}+C_{22}-\displaystyle\frac{C_{11}C_{22}-C_{12}^{2}}{C_{66}}\right)\sin^2(\theta)\cos^2(\theta)}{C_{11}\sin^4(\theta)+C_{22}\cos^4(\theta)-\left(2C_{12}-\displaystyle\frac{C_{11}C_{22}-C_{12}^{2}}{C_{66}}\right)\sin^2(\theta)\cos^2(\theta)},
\end{equation}
where $\theta$ represents the angular dependence of Young's modulus and Poisson's ratio.

\begin{figure}[t] 
  \centering \includegraphics[clip,width=7cm]{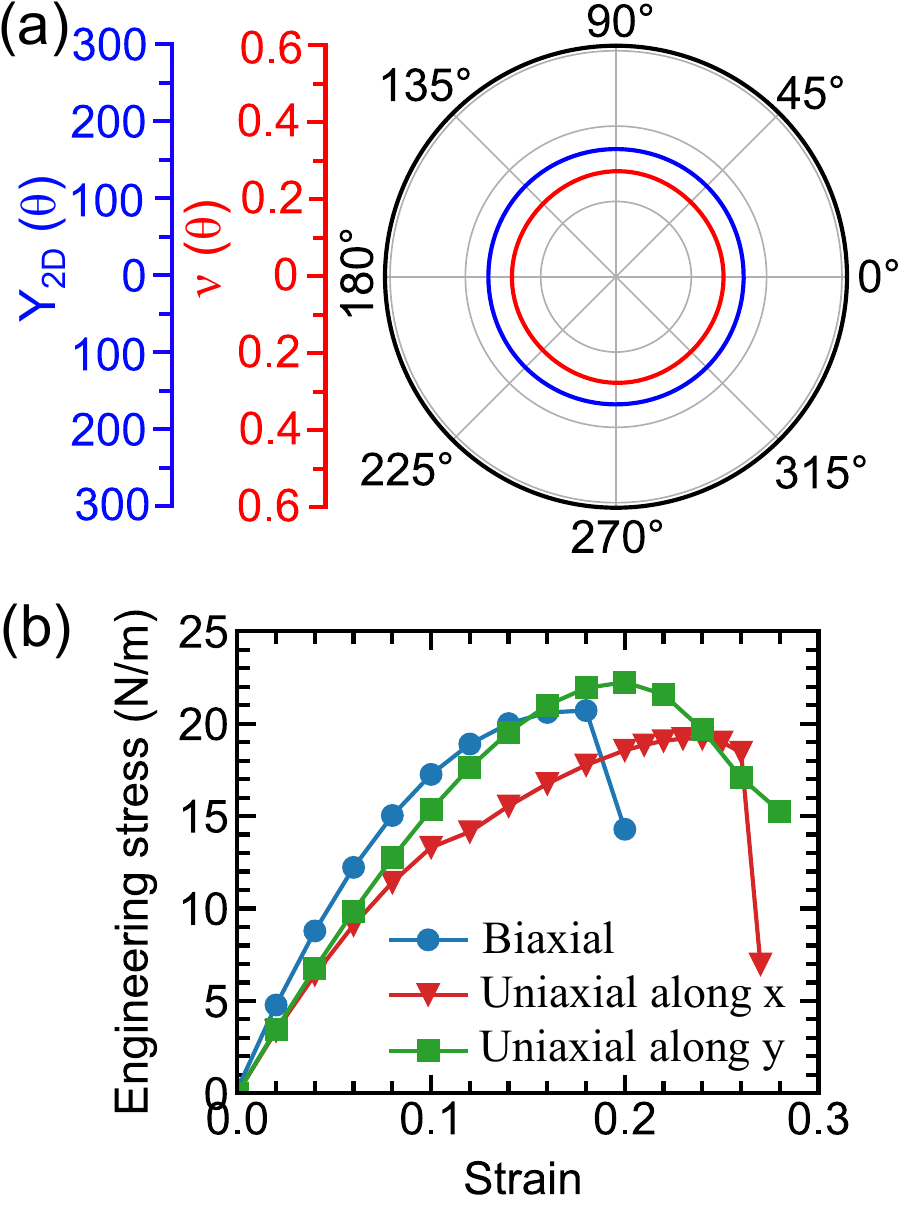}
  \caption{(a) Young's modulus $Y_{\text{2D}}(\theta)$ and Poisson's ratio ($\upsilon(\theta)$) as a function of the angle $\theta$. (b) Stress-strain curves along the biaxial strain and uniaxial strains along the $x$- and $y$-directions.}
  \label{fig:mechanical}
\end{figure}

In Figure~\ref{fig:mechanical}(a), we plot Young's modulus and Poisson's ratio as a function of the angle $\theta$, respectively. Our calculations indicate that both quantities exhibit isotropic behavior. The calculated $Y_{\text{2D}}$ of MoTe$_{2}$/PtS$_{2}$ is 170 N/m, which is higher than that of the PtS$_{2}$ monolayer (80.59 N/m)~\cite{du2018elastic} but lower than that of the MoTe$_{2}$ monolayer (116 N/m)~\cite{van2018charge}. Similarly, the Poisson’s ratio ($\upsilon$) of 0.28 lies between the values for PtS$_{2}$ (0.27)~\cite{du2018elastic} and MoTe$_{2}$ (0.32)~\cite{van2018charge}.

In Figure~\ref{fig:mechanical}(b), we plot the stress-strain curves of MoTe$_{2}$/PtS$_{2}$ under biaxial strain and uniaxial strains along $x$- and $y$-directions. The maximum stress and its corresponding strain are known as the ideal strength ($\sigma^i$) and critical strain $\varepsilon^c$, respectively, representing the mechanical strength~\cite{van2024janus, Thanh2023,li2012ideal,hung2023nonlinear}. For biaxial strain, $\sigma^i$ and $\varepsilon^{c}$ are 20.71 N/m and 0.18, repectively. The mechanical strength of MoTe$_{2}$/PtS$_{2}$ under biaxial strain are smaller than that of the AA-stacked MoSSe$_{2}$/MoS$_{2}$ (23.4 N/m, 0.21) and AB-stacked MoSSe$_{2}$/MoS$_{2}$ (23.1 N/m, 0.19)~\cite{hung2023nonlinear}, respectively. For uniaxial strains, $\sigma^i$ is 19.20 N/m and 22.25 N/m for the $x$- and $y$-directions, corresponding to $\varepsilon^c$ of 0.24 and 0.20, respectively. The critical strain, varying between $0.20$ and $0.24$ (or $20\% - 24\%$), highlights the suitability of MoTe$_{2}$/PtS$_{2}$ for use in flexible TE devices. In the next section, we will focus on investigating the effect of the biaxial strain on the TE properties. In the experiment, biaxial strain can be achieved using various techniques, such as external pressure or atomic force microscopy tip~\cite{dai2019strain}.

\subsection{Strain effect on electron and phonon properties}
\begin{figure*}[t]
  \centering \includegraphics[clip,width=15cm]{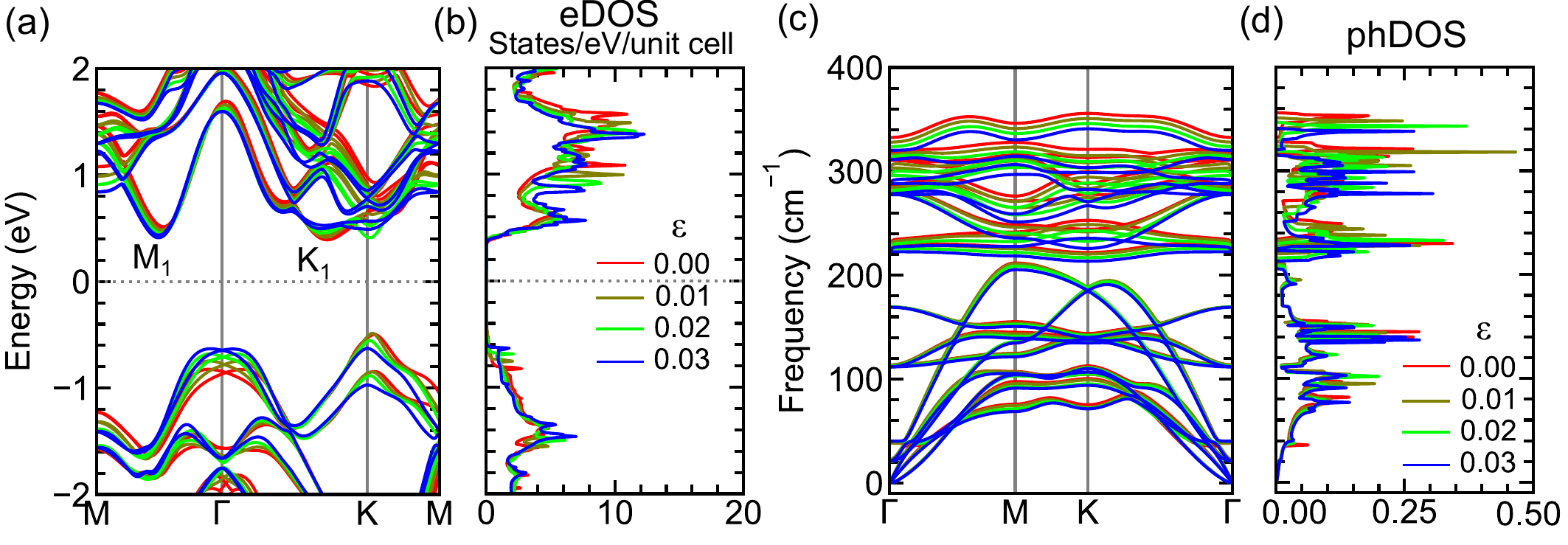}
  \caption{(a) Energy dispersion, (b) electronic density of states (eDOS), (c) phonon dispersion, and (d) phonon density of states (phDOS) of the 2D vdW heterobilayer MoTe$_{2}$/PtS$_{2}$ under the several values of biaxial strain $\varepsilon$.}
\label{fig:band-phonon-strain}
\end{figure*}

In Figures~\ref{fig:band-phonon-strain}(a) and (b), we plot the energy band structures and eDOS of MoTe$_{2}$/PtS$_{2}$ under several values of biaxial strain $\varepsilon$, respectively. The calculated results show that the CBM at the K$_{1}$ point shifts upward, while the conduction band at the M$_{1}$ point shifts downward, as shown in Figure~\ref{fig:band-phonon-strain}(a). In contrast, the VBM at the K point moves downward, whereas the valence band at the $\Gamma$ point shifts upward, potentially enhancing band convergence. For n-type MoTe$_{2}$/PtS$_{2}$, band convergence is achieved at $\varepsilon = 0.02$ with the different energy between valley bands $\Delta E \approx 0$ eV, while for p-type MoTe$_{2}$/PtS$_{2}$, it occurs at $\varepsilon = 0.03$. This strain-induced band convergence can significantly improve the TE performance of MoTe$_{2}$/PtS$_{2}$.

In Figures~\ref{fig:band-phonon-strain}(c) and (d), we plot the phonon dispersion and phonon density of states (phDOS) of MoTe$_{2}$/PtS$_{2}$ under several values of $\varepsilon$, respectively.
We find that there are no negative phonon frequencies, indicating that MoTe$_{2}$/PtS$_{2}$ is dynamically stable under biaxial strain. $\varepsilon$ in the range of $0.00-0.03$ almost does not change the acoustic phonon branches but leads to a slight reduction in the frequencies of the optical phonon branches. The optical phonon modes shift to lower frequencies, leading to a reduced phonon gap between the optical and acoustic modes. The frequencies of the LO1 and LO2 branches under strain can be found in Table S2 (Supporting Information) of the Supporting Information for reference. This suggests that the $\kappa_\text{ph}$ of MoTe$_{2}$/PtS$_{2}$ may not be significantly affected within this strain range. 
However, the biaxial strain can optimize band convergence, which could enhance the Seebeck coefficient and thereby improve the PF, as shown in Section 3.5.

\subsection{Strain effect on thermoelectric power factor}

\begin{figure*}[t] 
  \centering \includegraphics[clip,width=14cm]{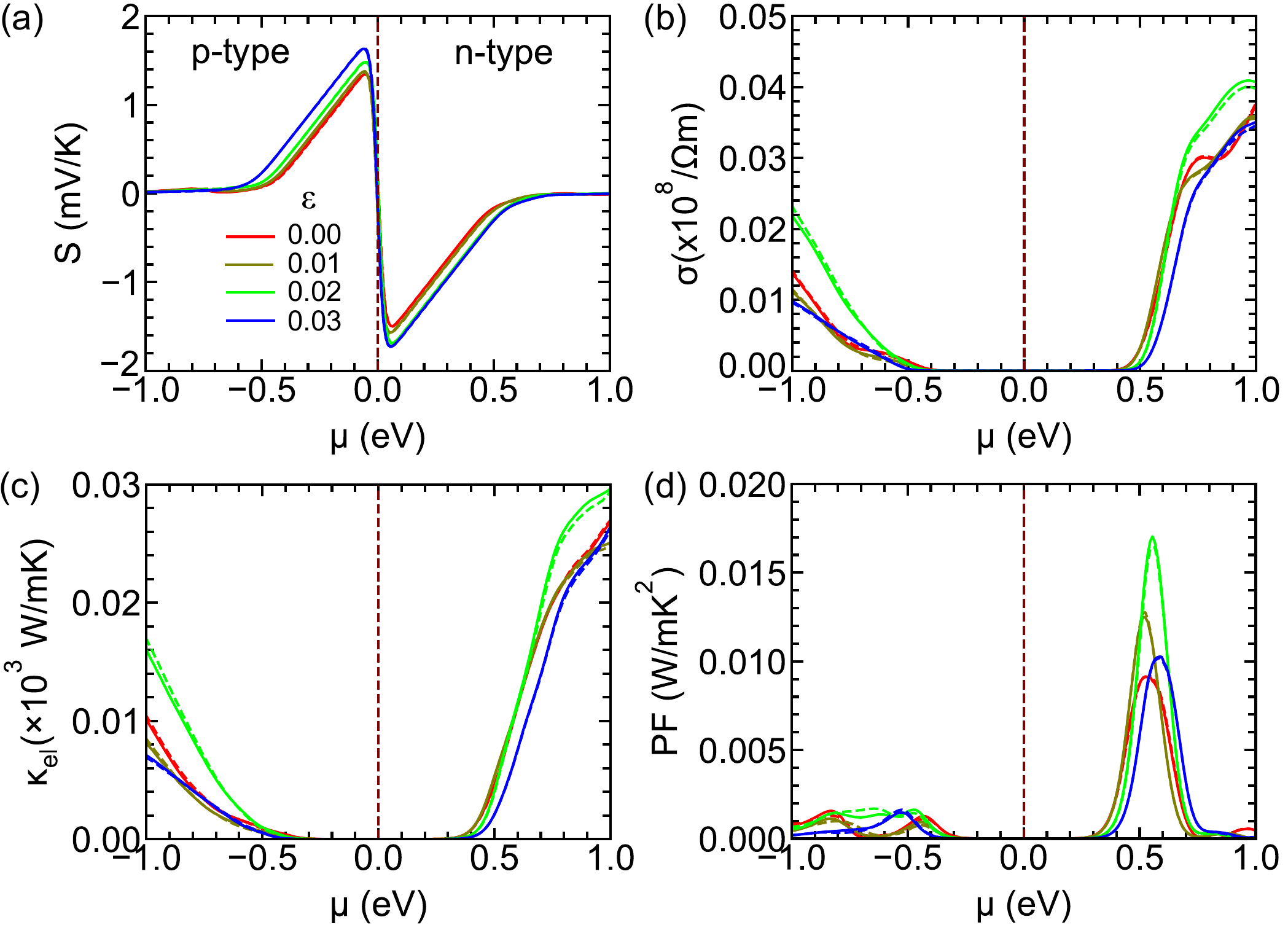}
  \caption{(a) Seebeck coefficient $S$, (b) electrical conductivity $\sigma$, (c) electronic thermal conductivity $\kappa_\text{el}$, and (d) power factor PF are plotted as a function of the chemical potential $\mu$ at several values of biaxial strain $\varepsilon$ for the p-type and n-type MoTe$_{2}$/PtS$_{2}$ at room temperature $T=300$ K.}
\label{fig:PF}
\end{figure*}

In Figure~\ref{fig:PF}(a), we show $S$ as a function of chemical potential $\mu$ for both p-type and n-type MoTe$_{2}$/PtS$_{2}$ at $T$ = 300 K under the biaxial strain $\varepsilon$. We find that $S$ increases with increasing strain in both p-type and n-type materials. It is noted that a high $S$ is often associated with a large density-of-states effective mass $m_{d}^{*}$, which is defined as $m_{d}^{*} = N_{v}^{2/3}m^{*}$~\cite{hung2019thermoelectric, lv2014enhanced}, where $N_{v}$ represents the number of degenerate band valleys. When the energy bands converge, $N_{v}$ increases, leading to a larger $m_{d}^{*}$ and, consequently, to an enhanced $S$. In the present study, band convergence is achieved for both the n-type and p-type by applying biaxial strain, as shown in Figure~\ref{fig:band-phonon-strain}(a). Without strain ($\varepsilon = 0.00$), the peak value of $S$ for MoTe$_{2}$/PtS$_{2}$ can be more than 1500 $\mu$V/K, which is higher than that of 2D vdW heterobilayer P/GeC (1210 $\mu$V/K)~\cite{ahmad2025van}, heterobilayer GeS/SiSe (1166 $\mu$V/K)~\cite{noshin2021modeling} and heterobilayer Sb/As (550 $\mu$V/K)~\cite{tang2022designing}. 
Additionally, we also calculated $S$ of MoTe$_{2}$/PtS$_{2}$ for several temperatures (see in Figure S7(a) in Supporting Information). In Figures~\ref{fig:TE}(b) and (c), we show the electrical conductivity $\sigma$ and the thermal conductivities by electron $\kappa_\text{el}$ as functions of $\mu$ for p-type and n-type materials. We note that the $\sigma$ and $\kappa_\text{el}$ values are scaled by $h/d_{0}$, where $h$ = 30 {\AA} and $d_{0}$ = 13.452 {\AA} represent the unit cell thickness and the effective material thickness, respectively. It is noted that to calculate $\sigma$ and $\kappa_\text{el}$, the relaxation time is also calculated for each $\varepsilon$, as listed in Table S3 (Supporting Information).

In Figure~\ref{fig:PF}(d), we show the PF for p-type and n-type MoTe$_{2}$/PtS$_{2}$ as a function of $\mu$ at $T$ = 300 K under the biaxial strain $\varepsilon$. Our results reveal that the n-type material achieves a significantly higher maximum PF compared to its p-type counterpart. This higher PF in the n-type material can be attributed to larger eDOS at near CBM compared with VBM, as shown in Figure~\ref{fig:band-phonon-strain}(b). At $\varepsilon =0.00 $, the optimum PF (PF$_{\text{opt}}$) of the n-type and p-type materials are \SI{92}{\micro\watt}/cmK$^{2}$ and \SI{16}{\micro\watt}/cmK$^{2}$, respectively. The PF$_{\text{opt}}$ of the n-type is higher than that of other materials, such as the 2D vdW heterobilayer Sb/As (\SI{40}{\micro\watt}/cmK$^{2}$)~\cite{tang2022designing} and CoSb$_{3}$ (\SI{40}{\micro\watt}/cmK$^{2}$)~\cite{shi2011multiple}. The maximum of PF$_{\text{opt}}$ for the n-type is found at $\varepsilon=0.02$ (or 2\%) because of the band convergence. The PF$_{\text{opt}}$ for the n-type with strain (\SI{170}{\micro\watt}/cmK$^{2}$ at $\varepsilon=0.02$) is enhanced by nearly 84.78\% of that without strain (\SI{92}{\micro\watt}/cmK$^{2}$ at $\varepsilon=0.00$). For p-type, at $\varepsilon = 0.00$, the PF$_{\text{opt}}$ exhibits two peaks at optimized chemical potential $\mu_{\text{opt}} = -0.44$ and $-0.83$ eV, in which the peak at $\mu_{\text{opt}} = -0.83$ eV originates from the contribution of the Rashba SOC at the $\Gamma$ point, as shown in Figure~\ref{fig:band-phonon-strain}(a). Under biaxial strain, these two peaks degenerate into one peak at $\mu_{\text{opt}} = -0.53$ at $\varepsilon=0.03$ due to the band convergence of the p-type case. In addition, $S$, $\sigma$, $\kappa_{el}$, and PF at different temperatures $T$ = 500, 700, and 900 K under biaxial strain are also calculated and shown in Figures S9, S10, and S11 in Supporting Information, respectively.  

\subsection{Strain effect on lattice thermal conductivity}

\begin{figure*}[t] 
  \centering \includegraphics[clip,width=14cm]{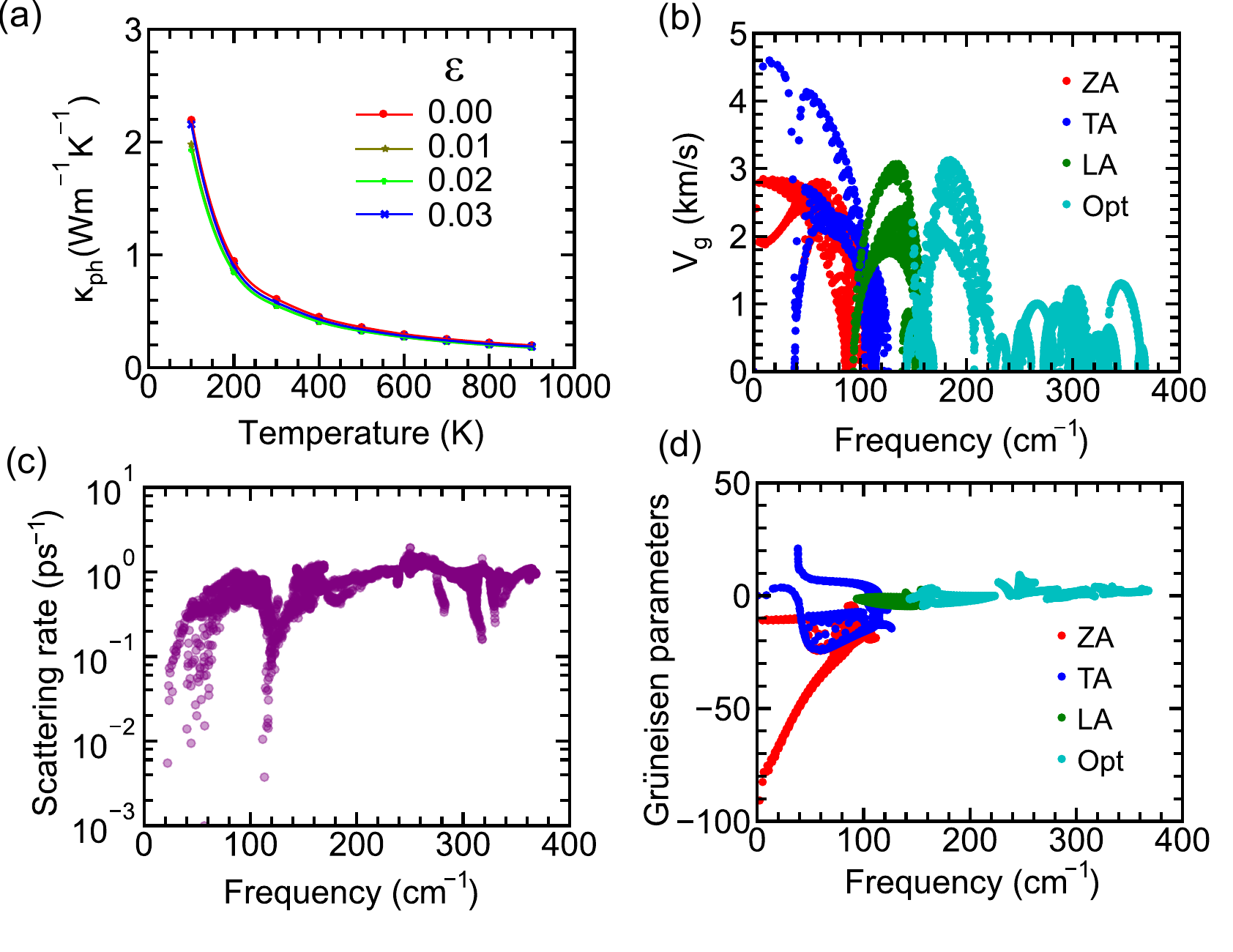}
  \caption{(a) Lattice thermal conductivity $\kappa_\text{ph}$ as a function of temperature under several values of biaxial strain $\varepsilon$, (b) phonon group velocity $v_g$, (c) phonon-phonon scattering rate, and (d) Gr{\"u}neisen parameter at $T$= 300 K of the 2D vdW heterobilayer MoTe$_{2}$/PtS$_{2}$.}
\label{fig:TE}
\end{figure*}

In Figure~\ref{fig:TE}(a), we show the lattice thermal conductivity $\kappa_\text{ph}$ of MoTe$_{2}$/PtS$_{2}$ as a function of temperature under several biaxial strains. We note that $\kappa_\text{ph}$ is scaled by a factor of $h/d_{0}$. We find that $\kappa_\text{ph}$ exhibits isotropic behavior and almost does not change with increasing strain from $\varepsilon=0.00$ to $\varepsilon=0.03$. On the other hand, $\kappa_\text{ph}$ decreases rapidly in the temperature range from 100 to 500 K. At temperatures above 500 K, the decreasing rate of $\kappa_\text{ph}$ slows down and tends to converge as the temperature increases. This phenomenon mainly originates from the intrinsic enhancement of phonon-phonon scattering because of the Umklapp process and $\kappa_\text{ph}$ $\propto 1/T$~\cite{zhu2019high,ahmad2025van}. At $T$ = 300 K, the calculated values of $\kappa_\text{ph}$ is 0.6 and 0.56 Wm$^{-1}$K$^{-1}$ at $\varepsilon=0.00$ and $0.03$, respectively. These values are lower than those of the well-known TE materials, such as Bi$_2$Te$_3$ and Mg$_3$Bi$_2$ ($\kappa_\text{ph}\sim 1-2$ Wm$^{-1}$K$^{-1}$)~\cite{hung2019designing,hung2023role}. 
Thus, MoTe$_{2}$/PtS$_{2}$ with extremely low $\kappa_\text{ph}$ can play a crucial role in achieving high $ZT$ value. 

The origin low $\kappa_\text{ph}$ of MoTe$_{2}$/PtS$_{2}$ can be attributed to the avoided crossing points between the acoustic phonon and optical phonon, which occur below 40 cm$^{-1}$, as shown in Figure~\ref{fig:k_pp-phonon}. Furthermore, the Bader charge-based ionic charge states of MoTe$_{2}$/PtS$_{2}$ are Mo$^{0.559+}$, Te$_{1}$$^{0.234-}$, Te$_{2}$$^{0.289-}$, Pt$^{0.458+}$, S$_{1}$$^{0.235-}$, and S$_{2}$$^{0.259-}$, indicating that Mo-Te$_1$, Mo-Te$_2$, Pt-S$_1$, and Pt-S$_2$ form metavalent bonds. The presence of metavalent bonding leads to strong anharmonicity and a short phonon relaxation time, as observed in the Janus $\gamma$-Ge$_{2}$SSe~\cite{Thanh2023} and B$_{2}$P$_{6}$~\cite{van2024janus} monolayers. As shown in Figure~\ref{fig:TE}(c), MoTe$_{2}$/PtS$_{2}$ shows short phonon relaxation time of $0.01-1.0$ ps$^{-1}$, indicating strong phonon scattering, consequently, a low $\kappa_\text{ph}$. 


Another factor that contributes to reducing $\kappa_\text{ph}$ is the phonon group velocity. As shown in Figure~\ref{fig:TE}(b), the calculated group velocity of the ZA and TA modes is about $1.9-4.6$ km/s. The acoustic phonon branches, which exhibit higher group velocities compared to the optical branches, mainly contribute to $\kappa_\text{ph}$.  To more understand the low lattice thermal conductivity, we calculate the Gr{\"u}neisen parameter $\left |\gamma \right |$ to measure the strength of anharmonicity of MoTe$_{2}$/PtS$_{2}$. The calculated results show that the highest $\left |\gamma \right |$ value of about 80 appears in the ZA branch, as shown in Figure~\ref{fig:TE}(d), which would greatly suppress the phonon transport in the high-temperature region. 

\subsection{Strain effect on thermoelectric figure of merit}

\begin{figure*}[t] 
  \centering \includegraphics[clip,width=14cm]{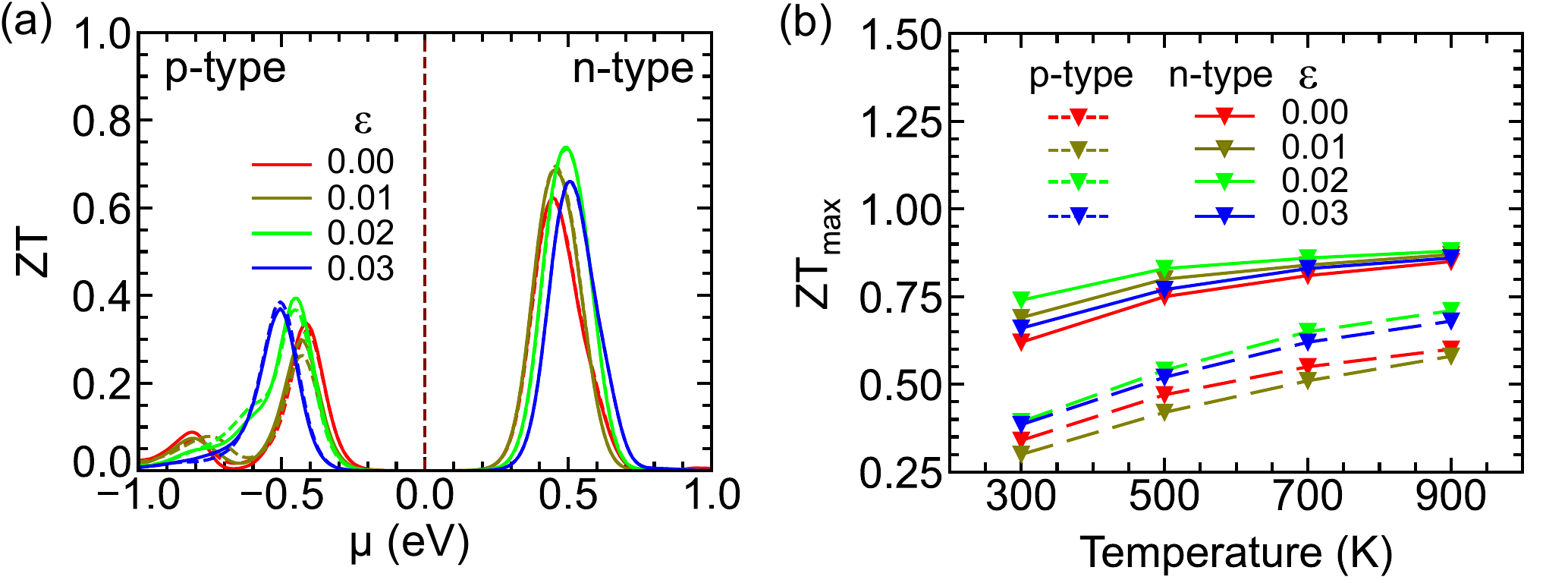}
  \caption{(a) Figure of merit $ZT$ is plotted as a function of $\mu$ at several values of biaxial strain $\varepsilon$ for p-type and n-type MoTe$_{2}$/PtS$_{2}$ at room temperature $T=300$ K. (b) The maximum value of $ZT$ as a function of temperature for p-type and n-type MoTe$_{2}$/PtS$_{2}$ at several $\varepsilon$ values.}
\label{fig:ZT}
\end{figure*}

In Figure~\ref{fig:ZT}(a), we show $ZT$ as a function of $\mu$ at $T=300$ for several values of $\varepsilon$. The maximum values of $ZT$ are about 0.40 and 0.74 for p-type and n-type, respectively. These $ZT$ values are comparable with that of the Janus Rashba monolayer Si$_{2}$SbBi (0.31 at 500 K), Si$_{2}$AsSb (0.38 at 500 K)~\cite{xia2024thermoelectric}, Rashba monolayer of WSTe (0.55 at 500 K)~\cite{duan2021room}, and 2D vdW heterobilayer MoSe${_2}$/BAs (0.58 at 1200 K)~\cite{li2021electronic}, and well-known TE materials ($ZT$ value typically ranges from 0.8 to 1.0 at 300 K). The $ZT$ is not significantly affected by the strain, although the PF value of n-type is significantly increased by band convergence. This is because at the optimum $\mu$ value of $ZT$ for n-type ($\mu_\text{opt}^{ZT}\sim 0.45$ eV), which is smaller than that of the PF for n-type ($\mu_\text{opt}^\text{PF}\sim 0.55$ eV), the PF is not significantly changed by the band convergence, as shown in Figure~\ref{fig:PF}(d). 

In Figure~\ref{fig:ZT}(b), we show the maximum $ZT$ ($ZT_\text{max}$) as a function of temperature ranging from 300 to 900 K for $\varepsilon$ ranging from 0.00 to 0.03. $ZT_\text{max}$ are 0.88 and 0.71 for the n-type and p-type, respectively, at $T = 900$ K and $\varepsilon = 0.02$. Since $ZT_\text{max}$ of the n-type changes about $19-37$\% for the temperature ranging from 300 to 900 K, it suggests that 2D vdW heterobilayer MoTe$_{2}$/PtS$_{2}$ could be a good candidate material for the TE application in the mid-range temperature.

\section{Conclusions}
In summary, we have investigated the effect of strain on the mechanical, electronic, phonon, and thermoelectric properties of the 2D vdW heterobilayer MoTe$_{2}$/PtS$_{2}$. A strong Rashba splitting is observed in MoTe$_{2}$/PtS$_{2}$, and the critical strain of about 20- 24\% suggests that MoTe$_{2}$/PtS$_{2}$ can be utilized in flexible TE devices. The Fr\"ohlich interaction model is employed to evaluate carrier mobility, which indicates the significant contribution of longitudinal-optical-phonon scattering to the TE properties of MoTe$_{2}$/PtS$_{2}$. We find that at a strain of $\varepsilon = 0.02$, the PF of the n-type material is significantly enhanced, reaching a high value of \SI{170}{\micro\watt}/cmK$^{2}$ W/mK$^{2}$ at 300 K, due to band convergence. The optimum $ZT$ values for n-type and p-type materials reach 0.88 and 0.71 at 900 K, respectively. Our calculated results demonstrate that the 2D vdW heterobilayer MoTe$_{2}$/PtS$_{2}$ holds potential as a flexible TE material in the mid-range temperature. 


\section*{Supporting Information}
The Supporting Information is available free of charge at \url{https://pubs.acs.org/doi/xxx.}

Table of Born effective charges, frequencies of the LO1 and LO2 phonon branches. Figure of the total energy and temperature as a function of time at $T$ = 900 K, figure of the band structure of MoTe$_{2}$/PtS$_{2}$ heterobilayer ($\sqrt{3} \times 1$ supercell), figure of the total energy-strain relationship, figure of the calculated band energies, figure of the temperature-dependent relaxation time at several strain values, figure of the Seebeck coefficient $S$, electrical conductivity $\sigma$, electronic thermal conductivity $\kappa_{el}$, and power factor PF are plotted as function of the chemical potential $\mu$ at $T$= 500 K, 700 K and 900 K.

\section*{Acknowledgements}
N.T.H. acknowledges the financial support from Frontier Research Institute for Interdisciplinary Sciences, Tohoku University.

\providecommand{\latin}[1]{#1}
\makeatletter
\providecommand{\doi}
  {\begingroup\let\do\@makeother\dospecials
  \catcode`\{=1 \catcode`\}=2 \doi@aux}
\providecommand{\doi@aux}[1]{\endgroup\texttt{#1}}
\makeatother
\providecommand*\mcitethebibliography{\thebibliography}
\csname @ifundefined\endcsname{endmcitethebibliography}  {\let\endmcitethebibliography\endthebibliography}{}

\end{document}